\documentclass[aps,prl,twocolumn,superscriptaddress,10pt]{revtex4-2}
\usepackage[utf8]{inputenc}
\usepackage{amsmath, bm}
\usepackage{cancel}
\usepackage{soul}
\usepackage{amsfonts}
\usepackage[normalem]{ulem}
\usepackage{mathtools}
\usepackage{physics}
\usepackage{dsfont}
\usepackage{verbatim}
\usepackage{color}
\usepackage[dvipsnames]{xcolor}
\usepackage[ruled,longend]{algorithm2e}
\usepackage{textgreek}
\usepackage{amssymb}
\usepackage{eucal}
\usepackage{tcolorbox}
\usepackage{wasysym}
\usepackage{float}
\usepackage{subfigure}
\usepackage{amsthm, amssymb,bbm}
\def\Var{\mathop{\rm Var}} 

\begin{document}
\title{Exponential concentration and symmetries in Quantum Reservoir Computing} 

\author{Antonio Sannia}
\email{sannia@ifisc.uib-csic.es}
\affiliation{%
 Institute for Cross-Disciplinary Physics and Complex Systems (IFISC) UIB-CSIC, Campus Universitat Illes Balears, 07122, Palma de Mallorca, Spain.
}%

\author{Gian Luca Giorgi}
\affiliation{%
 Institute for Cross-Disciplinary Physics and Complex Systems (IFISC) UIB-CSIC, Campus Universitat Illes Balears, 07122, Palma de Mallorca, Spain.
}%

\author{Roberta Zambrini}%
\affiliation{%
 Institute for Cross-Disciplinary Physics and Complex Systems (IFISC) UIB-CSIC, Campus Universitat Illes Balears, 07122, Palma de Mallorca, Spain.
}%

\begin{abstract}

Quantum reservoir computing (QRC) is an emerging framework for near-term quantum machine learning that offers in-memory processing, platform versatility across analogue and digital systems, and avoids typical trainability challenges such as barren plateaus and local minima. The exponential number of independent features of quantum reservoirs opens the way to a potential performance improvement compared to classical settings. However, this exponential scaling can be hindered by exponential concentration, where finite-ensemble noise in quantum measurements requires exponentially many samples to extract meaningful outputs, a common issue in quantum machine learning.
In this work, we go beyond static quantum machine learning tasks and address concentration in QRC for time-series processing using quantum-scrambling reservoirs. 
Beyond discussing how concentration effects can constrain QRC performance, we demonstrate that leveraging Hamiltonian symmetries significantly suppresses concentration, enabling robust and scalable QRC implementations. We illustrate our approach with concrete examples, including an established QRC design.
 
\end{abstract}
\maketitle

\textit{- Introduction} Reservoir computing (RC) is a machine learning framework designed for time series analysis, originally introduced to address the trainability challenges of recurrent neural networks~\cite{BookRC}. The core concept involves leveraging the dynamics of a system to process input signals, with the training phase focusing solely on optimizing a linear combination of measured observables. This approach enables a single reservoir computer to handle diverse tasks without encountering trainability issues. RC has been implemented across various platforms, ranging from digital systems~\cite{Jaeger2004,Maass2004} to analog devices~\cite{VanderSande2017,Tanaka2019,Nakajima2020}.

Recently, quantum reservoir computing (QRC) has attracted significant attention for its potential to enhance information processing by harnessing the unique properties of quantum systems~\cite{Fujii2017,Mujal2021}. Platforms such as digital~\cite{Chen2020,Kobayashi2024} and analog~\cite{Senanian2024} quantum circuits, bosonic~\cite{Nokkala2021, Govia2021,GarcaBeni2023,Skin} and fermionic~\cite{Ghosh2019,Llodr2022} models, photonic integrated circuits~\cite{Spagnolo2022}, and spin networks~\cite{Fujii2017, MartnezPea2021,Tran2021,Mujal2023,Sannia2024} have been explored as quantum reservoirs. The quantum regime offers two key advantages: the ability to exploit exponentially larger degrees of freedom compared to classical systems and the capability to analyze quantum data without relying on full-state tomography.

While QRC is considered a promising candidate for overcoming the fundamental limitations of other algorithms based on Quantum Neural Networks (QNNs)—such as the exponential cost of training due to barren plateaus~\cite{McClean2018,reviewbarrenplateaus,Cerezo2021,Sannia2024_BP,Thanasilp2023} and local minima~\cite{Anschuetz2022}—it is not without challenges. Similar to quantum kernel methods~\cite{Thanasilp2024} and quantum extreme learning machines~\cite{xiong2024}, QRCs can suffer from exponential concentration of output values, inheriting this problem from static to temporal tasks. This means that as the system size increases, each output tends to become exponentially close to a fixed, input-independent value. In realistic scenarios with limited computational resources, this phenomenon can make the whole QRC algorithm inefficient, as the number of required measurements to obtain non-trivial outputs grows exponentially with system scaling.

In the ideal case, where an infinite number of measurements is available, the learning ability of QNNs has been linked to their capacity for scrambling information~\cite{Shen2020, Wu2021}. However, training QNNs, which function as scramblers, is resource-inefficient~\cite{OrtizMarrero2021, Garcia2022}. Similarly, QRC tends to achieve peak performance when reservoirs operate as information scramblers~\cite{MartnezPea2021,Kobayashi2024,Xia}, even in the time-independent case of extreme learning machines~\cite{Vetrano2025}. Beyond ideal conditions, when finite-ensemble noise in quantum measurements is accounted for, we show that exponential concentration effects emerge, limiting achievable efficiency in terms of resources.
On a positive note, we identify a strategy to sustain QRC scalability and show that concentration can be limited by incorporating symmetries~\cite{Meyer2023,Larocca2022} into the reservoir model, the presence of which mitigates exponential concentration effects and preserves the distinguishability of output observables.

\textit{Quantum reservoir model --} 
\begin{figure}[t]
    \centering  \includegraphics[width=\linewidth, keepaspectratio]{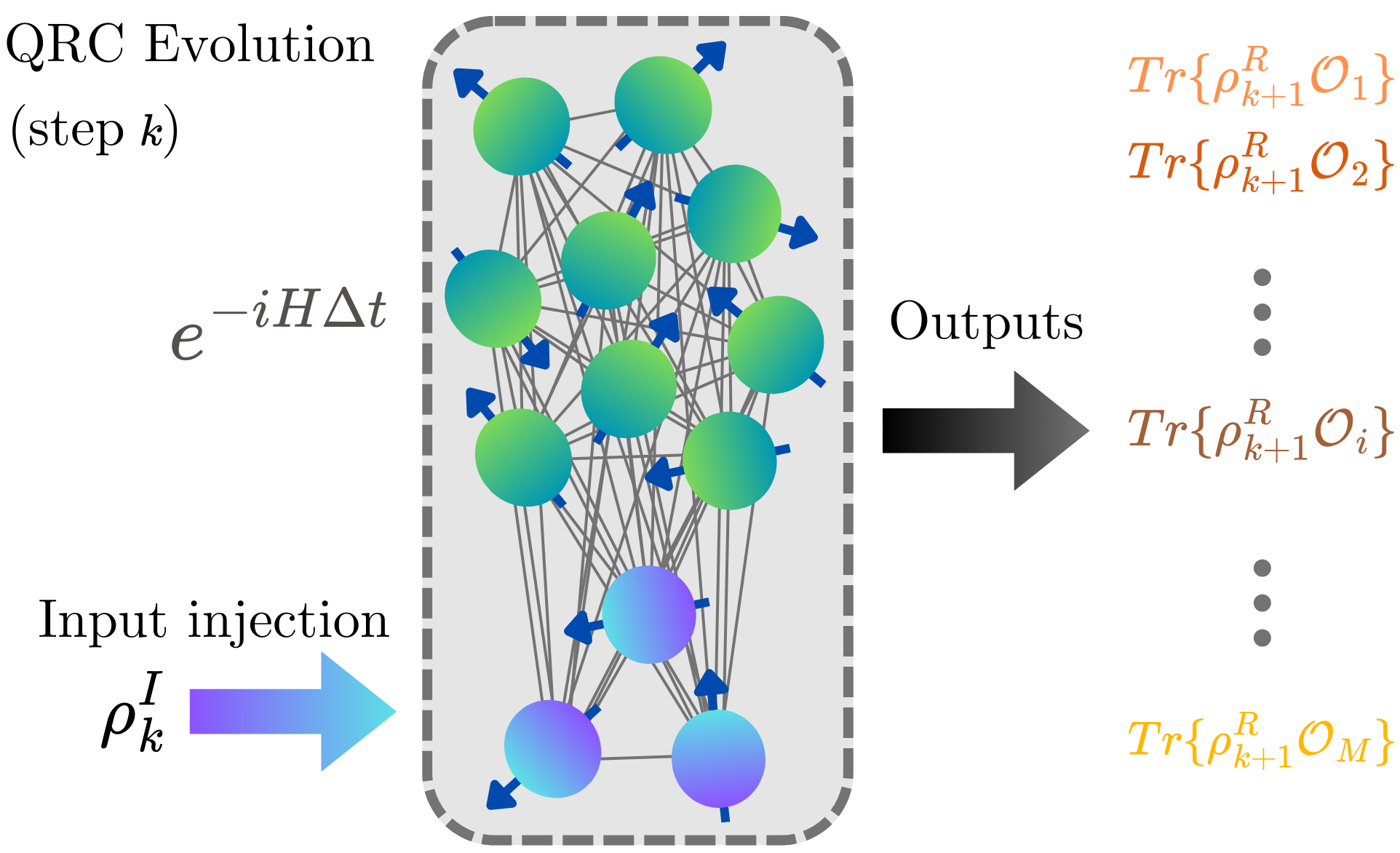}
    \caption{Three-layer design of QRC governed by Eq.~\ref{Eq:QRC}: Input injection in state $\rho_k^I$; reservoir evolution; output measurements (colors indicate that they are distinguishable in the ideal case without statistical errors).}
    \label{fig:Fig1}
\end{figure}
 We consider spin-based QRC models belonging to the class of erase-and-write maps, introduced in Ref.~\cite{Fujii2017}. These systems have been extensively studied and have shown promising results in processing both classical and quantum data series~\cite{Fujii2017, Tran2021}. The QRC algorithm can be summarized in three main steps: ($i$) input injection; ($ii$) dynamical evolution of the reservoir; ($iii$) measurement of reservoir observables to construct the output layer~\cite{Cucchi2022}. 

Consider a reservoir of $n$ qubits. At each step $k$ of the QRC algorithm, an input of $m \leq n$-qubit input states is injected $(i)$, resetting a subset of the reservoir qubits.   The entire reservoir then evolves according to its Hamiltonian, coupling the $n+m$ qubits. 
The reservoir updating rule $(ii)$ is expressed as:
\begin{equation}\label{Eq:QRC}
    \rho^R_{k+1} = e^{-iH\Delta t} \rho^{I}_k \otimes \Tr_{m}\{\rho^R_k\} e^{iH\Delta t},
\end{equation}
where $\Tr_m$ denotes the partial trace over the $m$ qubits into which the inputs are injected, $\rho^R_k$ is the reservoir state at time step $k$, $H$ is the reservoir Hamiltonian, and $\Delta t$ is the time between inputs (See Fig.~\ref{fig:Fig1} for a summary scheme). Notably, the input states $\rho^{I}_k$ may either encode classical data or arise from prior quantum processes, making the procedure suitable for both quantum and classical tasks \cite{Mujal2021}.

After each input injection, the reservoir outputs are derived from a set of $M$ observables $\{\mathcal{O}_i\}_{i=1}^{M}$. The output at time step $k$ is given by a linear combination:
\begin{equation}\label{Eq:output}
    y_k = \sum_{i=1}^{M}w_i \Tr{\rho_k \mathcal{O}_i} = \sum_{i=1}^M w_i \left\langle \mathcal{O}_i \right\rangle_k,
\end{equation}
where the $w_i$ values are the free weights optimized during training, and $\left\langle \mathcal{O}_i \right\rangle_k = \Tr{\rho_k \mathcal{O}_i}$. Minimal optimization resources are used to complete this final step ($iii$), typically through linear regression. 

The state space and the number of distinct observables scale exponentially with the full system size in a quantum systems.
In the ideal case, with unlimited resources, these expectation values can be computed exactly. However, in practice, only a finite number of measurements is available. This introduces unavoidable statistical errors that fundamentally limit the precision of reservoir output extraction~\cite{GarcaBeni2023, Mujal2023,Hu2023}. 
Moreover, for the expectation values in Eq.~\eqref{Eq:output} to be usable by the QRC algorithm, they must depend solely on the input series. This requirement necessitates that the reservoir dynamics satisfy the so-called \textit{echo state property}, which means that the reservoir state forgets its initial condition over time, making the output layer a well-defined function of the input series~\cite{YILDIZ20121}.

Not all dynamics described by Eq.~\eqref{Eq:QRC} satisfy the echo state property, and specific classes of reservoir Hamiltonians $H$ must be chosen. In particular, it has been found that the unitary evolution of the reservoir must scramble input-state information across all the system qubits to capture the desired feature~\cite{MartnezPea2021,Xia}.

\textit{Theory framework --} 
 Although the map in Eq.~\eqref{Eq:QRC} involves both unitary evolution and nonunitary state reset, analyzing its unitary part alone suffices to address the concentration problem, as this already emerges in each of the time steps. Assuming a chaotic reservoir Hamiltonian $H$, its unitary evolution $U = e^{-iH\Delta t}$ generates information scrambling when the evolution time $\Delta t$ exceeds a characteristic timescale. In this regime, $U$ can be regarded as a random unitary sampled from a 2-design distribution, satisfying what is known as ergodicity in classical statistical mechanics~\cite{Roberts2017}. This implies that, in the absence of Hamiltonian constraints, the state of the reservoir $\rho_k^R$ will converge exponentially  quickly (in the number of qubits $n$) toward the maximally mixed state, which renders it ineffective for information extraction. In this work, we propose to counteract this effect by identifying the proper class of scrambling reservoirs for QRC. We will consider block-diagonal Hamiltonians arising from the presence of symmetries in the system, which partition the Hamiltonian into invariant subspaces. This constrains the distribution from which $U$ is sampled, and, consequently, the resulting reservoir state distribution. The absence of symmetry can be seen as a special instance where the only conserved quantity, during the unitary step, is the energy. Our first result, proven in the Supplementary Material~\cite{sm}, characterizes these constrained reservoir states:

\textbf{Lemma 1}: \textit{Let $H$ be the reservoir Hamiltonian, and let $\{S_i\}_{i=1}^{K}$ denote the set of its corresponding irreducible symmetries, such that $[S_i, H] = 0$ and $[S_i, S_j] = 0$ for all $i,j$. If the reservoir unitary evolution $U = e^{-iH\Delta t}$ generates scrambling dynamics, then the reservoir state, at each time step $k$, can be described as being sampled from a probability distribution whose mean $\Bar{\rho}^R_{k}$ is given by the following direct sum:}
\begin{align}\label{Eq:Scramb_Res}
    \Bar{\rho}
    ^R_{k} = \bigoplus_{l=1}^{L} \alpha_l^{k} \frac{\mathbb{P}_l}{D_l}.
\end{align}
Here, $L$ is the number of irreducible symmetry sectors (joint eigenspaces of ${S_i}$), which we will call symmetric subspaces. If we denote by $d_i$ the number of distinct
eigenvalues of $S_i$, we have $L = \prod_{i=1}^{K} d_i$. The operators $\mathbb{P}_l$ are projectors onto these subspaces, and $D_l$ represents their corresponding dimension. Finally, the coefficients $\alpha_l^k$ depend on the particular input history in the QRC dynamical map (\ref{Eq:QRC}), as shown in the supplementary material~\cite{sm}.

By incorporating a variance analysis of the reservoir state distribution, in alignment with results found in other quantum machine learning models~\cite{Thanasilp2024, xiong2024}, we derive a condition that implies an exponential concentration for output observables. While previous analyses attribute this concentration to the exponential size of the entire Hilbert space, we anticipate that, in the presence of symmetries, it is a consequence of the possible exponential size of the symmetric subspaces. 

\textbf{Theorem 1} (Exponential concentration): \textit{Let $\mathcal{O}_i$ be an output observable. If, for all the symmetric subspaces, labelled by $l$, the following conditions hold: 
\begin{align*}
     \Tr{ \mathbb{P}_l \mathcal{O}_i} &= O(1); \\
    D_l &= O(e^{cn})
\end{align*}
where $c$ is a positive constant, then we have that 
\begin{equation*}
    \Tr\{\mathcal{O}_i \rho_k^R\} = O(e^{-cn}),
\end{equation*}
with probability exponentially close, with respect to $n$, to $1$. In the absence of symmetries, the same result holds, with $D_l$ dimension of the whole Hilbert space.}

Before entering into the importance of this theorem, we would like to add a result that follows from its demonstration in the Supplementary Material~\cite{sm} and is valid independently of the $D_l$ scaling. The expectation value $\Tr{\mathcal{O}_i \rho_k^R}$ is a sample from a probability distribution whose mean is the expectation value $\mathcal{O}_i$ computed with respect to the state of Eq.~\eqref{Eq:Scramb_Res} and the variance can still be evaluated asymptotically being  at most a $O(1/\min_{D_l \neq 1} D_l)$, also when not all the $D_l$ values scale exponentially. These considerations can be useful in addressing cases not affected by exponential concentration (see, for instance, the Ising model example below).
 
Under the hypothesis of Theorem  1, the presence of exponential concentration implies that a number of measurements that exponentially scale with the number of qubits is required to properly resolve the observable expectation values, rendering the reservoir computing algorithm inefficient. 
Nevertheless, Theorem 1 also provides clues for the strategies that allow one to circumvent concentration by violating its hypotheses. A first strategy could consist of considering reservoir Hamiltonians with a number of symmetries that scale with the size, ensuring that all the dimensions $D_l$ are at most $O(poly(n))$. In this case, the number of resources required to construct the output layer scales at most polynomially with the reservoir size, ensuring the overall efficiency of the algorithm.

As a second strategy, exponential concentration can be avoided by selecting observables that are compatible with the state structure of Eq.~\eqref{Eq:Scramb_Res}. More formally, we state

\textbf{Theorem 2} (Suitable observables): \textit{Given a reservoir dynamics that fulfill Lemma 1 hypothesis, let $\mathcal{O}_i$ be an output observable of the form:}
\begin{equation}\label{Eq:Obs_block}
    \mathcal{O}_i = \bigoplus_{l} \beta_l \mathbb{P}_l,
\end{equation}
\textit{which aligns with Eq.~\eqref{Eq:Scramb_Res}. If the amplitudes $\alpha_l^{k}$ and $\beta_l$ are independent of the system size, then $\Tr\{\mathcal{O}_i \rho_k^R\}$ can be considered as sampled from a probability distribution whose mean is $\sum_{l} \alpha_l^{k} \beta_l = O(1)$, avoiding an exponential concentration.}

Theorem 2, proven in the Supplementary Material~\cite{sm}, suggests that a natural choice of observables that avoids exponential concentration is the set of symmetry operators $S_i$ themselves. Furthermore, we observe that the number of degrees of freedom that can be efficiently extracted (related to the output layer) is determined by the number of functions $\alpha_l^{k}$, which is equivalent to the number of symmetric subspaces, $L$, in Eq.~\eqref{Eq:Scramb_Res}. 

Another important implication of Theorem 2 is that it establishes a necessary condition for the  effective operation of the reservoir computing model. Specifically, a dynamical reservoir response can only be achieved if the $\alpha_l^{k}$ coefficients vary at each step. This variation implies that the injection of input states, as described in Eq.~\eqref{Eq:QRC}, must change the expectation values of the symmetries with respect to the reservoir state. 

As in the examples we show below, the echo state property is directly satisfied when input injections constrain the coefficients $\alpha_l^{k}$ to be fading memory functions, which means that they only depend on the recent input history~\cite{Cucchi2022}. In this case, with good approximation, 
\begin{equation*}
    \alpha_l^{k} \simeq \alpha_l^{k}(\rho_{k-1}^I, \dots, \rho_{k-\tau_{FM}}^I),
\end{equation*}
where $\tau_{FM}$ is the fading memory time. It is now immediate to conclude that $\Tr\{\mathcal{O}_i \rho_k\}$ is a sample from a distribution that only depends on the input series. This distribution is independent of the initial state, thereby satisfying the echo state property requirement. 

{Importantly, Theorems 1 and 2 imply that for a scrambling reservoir, as in the case of Eq.~\eqref{Eq:QRC}, symmetries are necessary to prevent exponential concentration. Formally, we state:}

\textbf{Corollary 1} (Necessity of symmetries):  \textit{For a reservoir evolving according to a scrambling unitary, an efficient protocol to extract input-dependent dynamics, avoiding exponential concentration, must exhibit symmetries.}

In fact, in the absence of symmetries, the only way of avoiding an exponential concentration, violating Theorem 1 hypothesis, is to consider an observable of the form of Eq.~\eqref{Eq:Obs_block}, as dictated by Theorem 2. Then, only observables proportional to the identity satisfy this condition, yet their expectation values are obviously input-independent and useless for QRC. Consequently, symmetries become a necessary requirement for scalable reservoir computing, as formalized in Corollary 1.

\textit{Analytical QRC example avoiding concentration--} To illustrate the theoretical framework developed above, we now present a tailored example where we can construct analytically the proper observables for a simple task of series discrimination:

\textbf{Definition 1} (Series discrimination task): \textit{Consider an $n$-qubit reservoir described by Eq.~\eqref{Eq:QRC}, initialized in a random state. We inject one of two possible time series of single-qubit states into the reservoir: $\{|0\rangle\langle0|\}_k$ or $\{|1\rangle\langle1|\}_k$. The goal of the task is to distinguish between these two cases without directly measuring the input-encoded qubit.}

Considering the reservoir's unitary evolution as a scrambler, efficiently solving the task requires that the scrambler produces a response independent of the initial condition, thereby satisfying the echo state property, and transmits the input information across the reservoir qubits, preventing exponential concentration. As discussed, this condition can only be achieved in the presence of symmetries, and we will now identify a symmetry operator suited for this particular problem. More formally, we state:

\textbf{Proposition 1} (Efficient task resolution): \textit{For the discrimination task in Definition 1, when the reservoir implements unitary scrambling dynamics, {a model that is symmetric under the operator  $S = \sum_{i=1}^n \sigma_i^z$, can solve the task efficiently, both in required time steps and resources needed for output extraction.}}

A proof of Proposition 1 is provided in the Supplementary Material~\cite{sm}, where the reservoir evolution is analytically computed. Furthermore, for a reservoir model to solve a task efficiently, the mere presence of symmetries is insufficient, as they must be properly matched to the problem structure. To illustrate this, if we repeat the calculations from the proof of Proposition 1, using a reservoir model symmetric under $S' = \sum_{i=1}^n \sigma_i^x$, we find that the reservoir will converge to the same final state for both input series,  failing to discriminate between them.

\textit{QRC with Ising reservoir --} The developed framework establishes the limitations to scalable QRC due to concentration and, more importantly, strategies to overcome them. Beyond general considerations and synthetic models,  we now apply our theory to a widely used quantum reservoir model, the first proposed in the literature. Specifically, we consider the known analogue reservoir with Hamiltonian given by a fully transverse-field Ising model~\cite{Fujii2017, MartnezPea2021,Tran2021,Mujal2023,Xia}:
\begin{equation}\label{eq:ising}
    H = \sum_{i>j=1}^n J_{ij}\sigma_i^x\sigma_j^x + \frac{1}{2}\sum_{i=1}^n(h + h_i)\sigma_i^z,
\end{equation}
where $\sigma_i^a$ ($a = x,y,z$) are the Pauli matrices, $J_{ij}$ are the coupling strengths uniformly sampled from the set $[-J_s/2, J_s/2]$, $h$ is a uniform magnetic field, and $h_i$ is an on-site disorder randomly sampled from the interval $[-W,W]$. In the following, all the hyperparameters will be expressed in the $J_s$ units. The inputs of Eq.~\eqref{Eq:QRC} are single-qubit states parameterized by a classical real parameter: 
\begin{equation*}
    \rho^{I}_k = | \psi_{s_k} \rangle \langle \psi_{s_k}|,
\end{equation*}
where $| \psi_{s_k} \rangle = \sqrt{s_k} | 0 \rangle + \sqrt{1 - s_k} | 1 \rangle$ with $s_k$ a real number belonging to range $[0,1]$.

It has been shown that the model satisfies the echo state property, which is a necessary condition for a functioning reservoir computer, only within specific Hamiltonian regimes. The optimal operating conditions emerge in the ergodic phase, where the Hamiltonian's chaotic dynamics naturally induce scrambling behavior~\cite{MartnezPea2021}. Consequently, the hyperparameters $W$ and $h$ must be carefully selected to satisfy this condition. As recalled in the Supplementary Material~\cite{sm}, by fixing the hyperparameters $W$=$10^{-2}$ and $h$=$10^1$ we fall in this scenario, and, from now on, we maintain these values. 

\begin{figure}[t]
    \centering  \includegraphics[width=\linewidth, keepaspectratio]{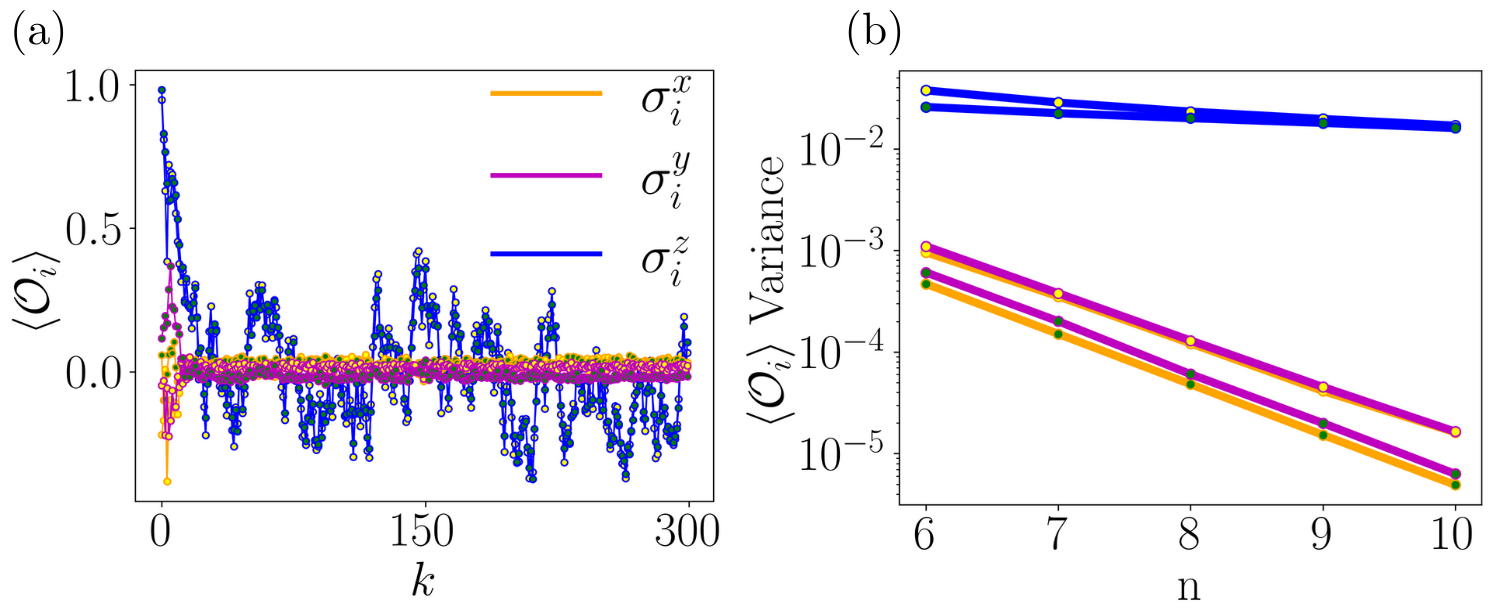}
    \caption{(a) Expectation values evolution of single-qubit Pauli observables, responding to a random input sequence for a 7-qubit realization. (b) Average variances of the expectation values series, over $1000$ random realizations of the Hamiltonian and initial reservoir states, varying the number of qubits, depicted in a logarithmic scale. $500$ random inputs have been used to wash out the dependence on the initial conditions, while with the following $500$, the variances have been computed. In the two plots, we represented the qubit on which the input is injected and a different qubit in the network, respectively with yellow and green points.}
    \label{fig:dyn}
\end{figure}

Consistent with our analytical results, only observables immune to exponential concentration are suitable for the construction of the output layer. This requires identifying the possible symmetries of the Hamiltonian $H$ of Eq.~\eqref{eq:ising}. In this particular case, the only exact symmetry is the parity $P = \prod_{i=1}^n \sigma_i^z$, which, despite being immune to scrambling effects, tends to vanish due to its global action on the qubit set~\cite{Cerezo2021,xiong2024}. 

However, in the deep thermal phase ($h\gg J_s$), the Hamiltonian is well-approximated by its leading-order term $H \simeq h/2\sum_{i=1}^n\sigma_i^z$. This dominant component identifies the operators that effectively approximate dynamical symmetries that can help identify a set of suitable observables, while the subleading contributions ensure the unitary evolution generates scrambling dynamics.
Under this approximation, the dimension of the symmetric subspaces is $D_l = \binom{n}{l}$. 
Let us first consider $l = c$ or $l = n - c$, with $c$ a constant independent of $n$. Although these subspaces are immune to exponential concentration as their size grows polynomially with $n$,  they display negligible populations (for a random input injection) and are not useful for computational purposes. Out of these negligible subspaces, concentration is expected to emerge from our theoretical analysis in all the remaining symmetry sectors due to their exponential dimension growth. In this case, Theorem 2 helps to find the observables that escape concentration, as for instance $\sigma_i^z$. The results presented in Fig.~\ref{fig:dyn} (a) confirm these considerations. In fact, only the expectation values of $\sigma_i^z$, which approximate the symmetries of the Hamiltonian, exhibit clearly distinguishable non-zero values, while $\sigma_i^x$ and $\sigma_i^y$ are very close to zero.

To quantify exponential concentration as a function of $n$, we computed the variance of the expectation value series. The averaged variances are depicted in Fig.~\ref{fig:dyn} (b). As expected, for the $\sigma_i^z$ observables, the variances persist at a system-size-independent order of magnitude, whereas for the $\sigma_i^x$ and $\sigma_i^y$ cases, they are exponentially suppressed as the system size increases. It implies that an output layer composed of single-qubit observables can be efficiently computed by considering only the $\sigma^z_i$ matrices. Still, the expectation values of different $\sigma_i^z$ tend to converge towards the same value. This is due to the permutation symmetry of the approximated Hamiltonian, and to the fact that Eq.~\eqref{Eq:Scramb_Res} is a good approximation of the reservoir state.

\begin{figure}[t!]
    \centering  \includegraphics[width=\linewidth, keepaspectratio]{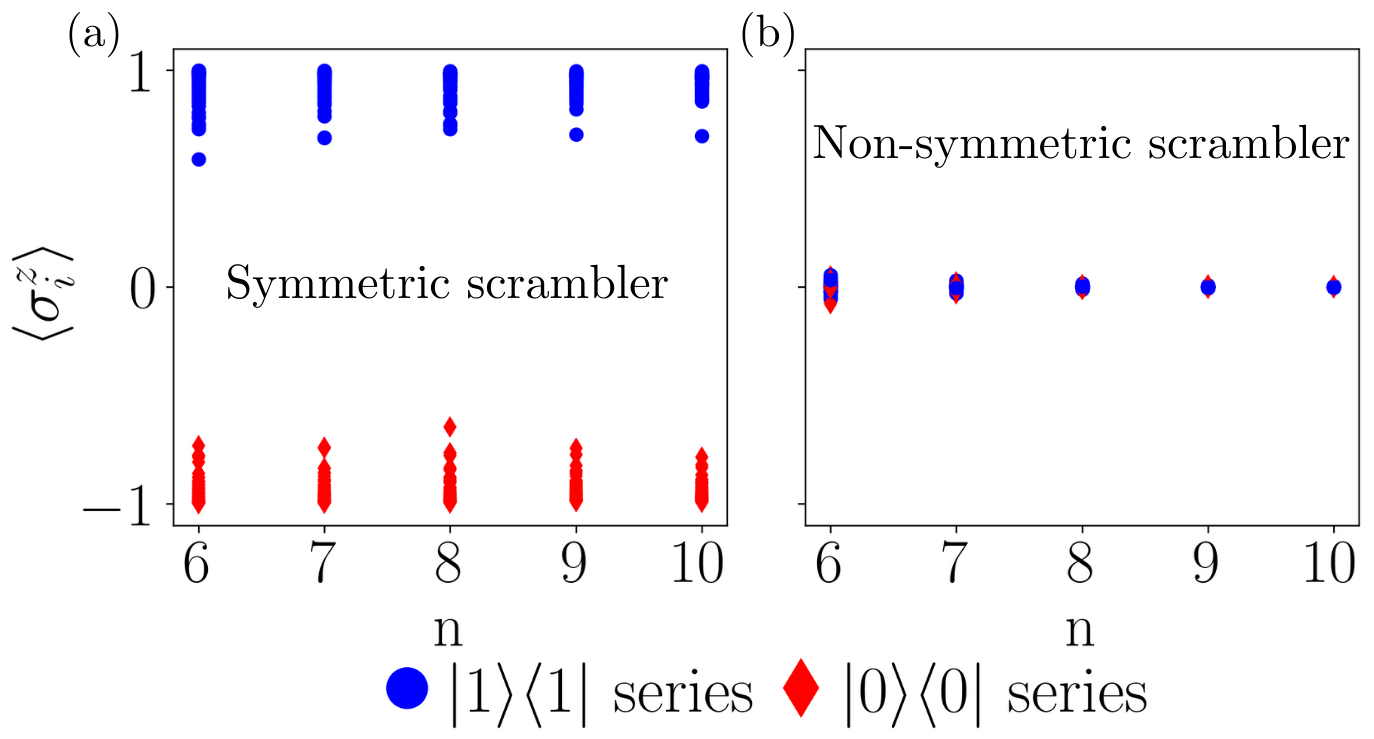}
    \caption{Numerical verification of Theorem 3. Reservoir response as a function of the number of qubits, $n$, for two cases: (a) Symmetric scrambler, generated by the Ising Hamiltonian in its thermal phase; (b) Asymmetric scrambler, obtained by sampling the reservoir unitary evolution from a Haar distribution. In both cases, 1000 input sequences of either $|1\rangle\langle1|$ or $|0\rangle\langle0|$ states were injected into random initial reservoir states. For a fixed $n$, 100 different realizations are shown for each input series. Each point represents the expectation value $\langle\sigma_i^z\rangle$, computed for a randomly chosen qubit distinct from the one receiving the input.}
    \label{fig:disc_tasl}
\end{figure}

Interestingly, this approximated $H$ naturally displays the symmetry of Proposition 1 and can therefore realize the series discrimination task of Definition 1. In Fig.~\ref{fig:disc_tasl}, we numerically calculate the expectation values of $\sigma_i^z$, where $i$ refers to a randomly chosen qubit distinct from the one where the input was applied. As predicted by our theoretical results, the Ising model of Eq.~\eqref{eq:ising}  (symmetric scrambler) can discriminate between the two cases, while this is not possible for a unitary that is randomly sampled from a Haar distribution (non-symmetric scrambler) that tends to concentrate.

\textit{Discussion and conclusion} We have demonstrated that exponential concentration effects pose a fundamental challenge to quantum reservoir computing, particularly when using scrambling dynamics. While scramblers efficiently propagate information across quantum systems, we demonstrate that this process exponentially suppresses the distinguishability of output observables. Interestingly, our results link the issues already known about learning scrambling dynamics to the possibility of using them as machine learning models~\cite{Holmes2021}. 

Crucially, we have shown that incorporating symmetries into the reservoir Hamiltonian prevents this exponential concentration, enabling scalable and efficient QRC. Our analytical results establish a direct connection between symmetries and the set of observables that remain distinguishable, while numerical simulations validate these predictions. 

While our analysis focuses on spin-based erase-and-write dynamics, the symmetry-enhanced framework is extensible to feedback-controlled systems~\cite{Kobayashi2024}, real-time learning protocols~\cite{Mujal2023,franceschetto2024}, hybrid classical-quantum architectures~\cite{Nokkala2024,Spagnolo2022,wudarski2024,settino2024}, and quantum extreme learning paradigms~\cite{Mujal2021}.

Our work establishes symmetry engineering as an indispensable paradigm for mitigating quantum reservoir computing’s inherent scalability limits. This opens the possibility of future research where optimal symmetry structures are systematically tailored for specific time series tasks of interest. 

Finally, we remark that, in addition to the resource requirements for extracting the output signal, another fundamental aspect of QRC is its memory behavior. In Ref.~\cite{sannia2025nonmarkovianity}, we demonstrated that avoiding an exponential memory decay, over time, necessitates the use of non-Markovian maps. Consequently, we highlight that symmetric non-Markovian reservoirs can exhibit long-range temporal correlations with the inputs, also enabling an efficient information extraction.

During the preparation of the manuscript, we became aware of a related independent work~\cite{Xiong_impact} that also investigates exponential concentration in quantum reservoir computing, but with a different focus. In our work, we propose symmetries as a strategy to solve exponential concentration effects that originate from the chaotic nature of the reservoir evolution. In contrast, in Ref.~\cite{Xiong_impact}, exponential concentration is studied as a function of the iterative application of the reservoir computing algorithm.

\section* {ACKNOWLEDGMENTS}
We acknowledge the Spanish State Research Agency, through the Mar\'ia de Maeztu project CEX2021-001164-M funded by the MCIU/AEI/10.13039/501100011033, through the COQUSY project PID2022-140506NB-C21 and -C22 funded by MCIU/AEI/10.13039/501100011033, MINECO through the QUANTUM SPAIN project, and EU through the RTRP - NextGenerationEU within the framework of the Digital Spain 2025 Agenda. The CSIC Interdisciplinary Thematic Platform (PTI+) on Quantum Technologies in Spain (QTEP+) is also acknowledged. The project that gave rise to these results received the support of a fellowship from the ``la Caixa” Foundation (ID 100010434). The fellowship code is LCF/BQ/DI23/11990081.

\bibliography{bibliography}

\clearpage
\setcounter{equation}{0}
\def\theequation{S\arabic{equation}}
\setcounter{figure}{0}
\onecolumngrid
\section*{Supplementary Material for "Exponential concentration and symmetries in quantum reservoir computing"}
\def\thefigure{S\arabic{figure}}

\section{Proofs of the analytical results presented in the main text}

\subsection{Proof of Lemma 1}
\textit{Proof:}

Given the commuting irreducible symmetry operators $\{S_i\}_{i=1}^{K}$ satisfying $[S_i, H] = 0$ and $[S_i, S_j] = 0$, we can define a basis of simultaneous eigenvectors, denoted as $|\psi^s_l\rangle$. Here, the index $l$, running up to $L$, identifies the joint eigenvalue sector of all $S_i$ (therefore, this index identifies each set of common eigenvectors of all symmetries).
The superindex $s$ instead identifies states within the 
$l$-th degenerate subspace (whose dimension is $D_l$). On this basis, the Hamiltonian $H$ can be expressed as a direct sum: 
\begin{equation*}
    H = \bigoplus_{l=1}^{L} H_l,
\end{equation*}
where each $H_l$ is a Hermitian operator that can be written as a linear combination of projectors of the form $|\psi^s_l\rangle\langle\psi^s_l|$.

Consequently, the reservoir unitary evolution becomes:
\begin{equation*}
    U = e^{-iH\Delta t} = \prod_{l=1}^{L} e^{-iH_l\Delta t} = \prod_{l=1}^{L} U_l,
\end{equation*}
where each $U_l$ is sampled from a 2-design distribution, consistent with the scrambling hypothesis~\cite{Roberts2017}.

Under this condition, using Haar integral identities, the distribution of $U_l$ satisfies the following superoperator equality~\cite{Puchaa2017,Mele2024}:
\begin{equation*}
    \int d\mu(U_l) \, U_l \left(\cdot\right) U_l^\dagger =
    \Tr\{\mathbb{P}_l \, \cdot\} \frac{\mathbb{P}_l}{D_l} +  \prod_{l'=1, l'\neq l}^{L} \mathbb{P}_{l'}\left(\cdot\right) \prod_{l'=1, l'\neq l}^{L} \mathbb{P}_{l'},
\end{equation*}
where with $d\mu(\cdot)$ we indicate a distribution volume element and $\mathbb{P}_l$ is the projector operator onto the eigenspace of $H_l$, defined as: $\mathbb{P}_l = \sum_{s=1}^{D_l} |\psi^s_l\rangle\langle\psi^s_l|$.

Using this identity, we can compute the average reservoir state after the unitary evolution:
\begin{equation*}
    \int d\mu(U) \, U \left(\rho_k^I \otimes \Tr_m\{\rho_k^R\}\right) U^\dagger = 
    \int \prod_{l} d\mu(U_l) \, U_L \cdots U_1 \rho_k^I \otimes \Tr_m\{\rho_k^R\} U_1^\dagger \cdots U_L^\dagger 
    = \bigoplus_{l=1}^{L} \alpha_l^k \frac{\mathbb{P}_l}{D_l},
\end{equation*}

where 
\begin{equation*}
    \alpha_l^k = \Tr\{\mathbb{P}_l \, \rho_k^I \otimes \Tr_m\{\rho_k^R\}\}.
\end{equation*}
$\blacksquare$

\subsection{Proof of Theorem 1}
\textit{Proof:}

From Eq.~\eqref{Eq:Scramb_Res} of Lemma 1, we find that $\Tr{\mathcal{O}_i \rho_k^{R}}$ is a value sampled from a probability distribution whose mean is:
\begin{equation*}
    \Tr{\mathcal{O}_i\Big(\bigoplus_{l=1}^{L} \alpha_l^{k} \frac{\mathbb{P}_l}{D_l}\Big)} = \sum_{l=1}^L \alpha_l^{k} \frac{\Tr{\mathcal{O}_i \mathbb{P}_l}}{D_l},
\end{equation*}
which is a $O(e^{-cn})$ under the Theorem hypotesis.

To conclude the proof of the theorem, it is now necessary to determine the variance of the distribution. Therefore, we need to evaluate the following quantity:
\begin{equation*}
    \Var_{U}[\Tr{\mathcal{O}_i \rho_k^R}] = \int d\mu({U}) \Tr{\mathcal{O}_i \rho_k^R}^2 - \left(\int d\mu({U}) \Tr{\mathcal{O}_i \rho_k^R} \right)^2.
\end{equation*}

As we already know that the second term in the sum is an $O(e^{-2cn})$, we proceed to evaluate the first one:
\begin{equation*}
    \int d\mu({U}) \Tr{\mathcal{O}_i \rho_k^R}^2 = \int d\mu({U}) \Tr{\mathcal{O}_i^{\otimes2} {\rho_k^R}^{\otimes2}} = \int \prod_l d\mu(U_l) \Tr{\mathcal{O}_i^{\otimes2} U_L^{\otimes2} \cdots U_1^{\otimes2} \tilde{\rho}_k^{\otimes 2} U_1^{\dagger \otimes2} \cdots U_L^{\dagger \otimes2}}.
\end{equation*}
where have used the property that the trace of the tensor product of two matrices is the product of their traces and the abbreviation $\tilde{\rho}_k = \rho^{I}_{k-1} \otimes \Tr_{m}\{ \rho^R_{k-1}\}$.

We now recall that, under the 2-design hypothesis, the second-order moments of the unitaries $U_l$ satisfy the following identity~\cite{Puchaa2017,Mele2024}:

\begin{equation*}
    \int d\mu(U_l) U_l^{\otimes2}(\cdot)U_l^{\dagger \otimes2} = \frac{\Tr\{ \mathbb{P}_{2l}\cdot\} - D_l^{-1}\Tr\{\mathbb{F}_{2l} \cdot \}}{D_l^2-1} \mathbb{P}_{2l} + \frac{\Tr\{\mathbb{F}_{2l} \cdot\} - D_l^{-1}\Tr\{\mathbb{P}_{2l}\cdot \}}{D_l^2-1} \mathbb{F}_{2l} + \prod_{l'=1, l'\neq l}^{L} \mathbb{P}_{2l'}(\cdot)\prod_{l'=1, l'\neq l}^{L} \mathbb{P}_{2l'},
\end{equation*}
where $\mathbb{P}_{2l}$ is the projector operator: $\mathbb{P}_{2l} = \sum_{s, s' = 1}^{D_l} | \psi^s_l \rangle | \psi^{s'}_l \rangle \langle \psi^s_l | \langle  \psi^{s'}_l |$, and
$\mathbb{F}_{2l}$ is the Flip operator: $\mathbb{F}_{2l} = \sum_{s, s' = 1}^{D_l} | \psi^s_l \rangle | \psi^{s'}_l \rangle \langle \psi^{s'}_l | \langle  \psi^s_l |$.

From a direct substitution of Eq.~\eqref{Eq:Scramb_Res}, we can easily find the leading order term of the desired quantity:
\begin{equation*}
     \int d\mu(U_l) \Tr{\mathcal{O}_i \rho_k^R}^2 \simeq \sum_{l=1}^{L} (\alpha_l^k)^2 \Tr{\mathbb{P}_l \mathcal{O}_i}^2/D_l^2 + \sum_{l=1}^{L}\Tr{\mathbb{P}_l\tilde{\rho}_k\mathbb{P}_l\tilde{\rho}_k}\Tr{\mathbb{P}_l\mathcal{O}_i\mathbb{P}_l\mathcal{O}_i}/D_l^2,
\end{equation*}
concluding that $\int d\mu(U_l) \Tr{\mathcal{O}_i \rho_k^R}^2 $ and, consequently, $\Var_{U}[\Tr{\mathcal{O}_i \rho_k^R}]$ are $O(e^{-2cn})$. At this point, Chebyshev’s inequality concludes the proof. $  \blacksquare$ 

\subsection{Proof of Theorem 2}
\textit{Proof:}

Combining Eqs.~\eqref{Eq:Scramb_Res} and~\eqref{Eq:Obs_block} of the main text, we find an expression for the $\Tr\{\mathcal{O}_i \rho_k^R\}$ mean:
\begin{equation*}
     \Tr{\Big(\bigoplus_{l=1}^{L} \alpha_l^{k} \frac{\mathbb{P}_l}{D_l}\Big) \Big(\bigoplus_{l} \beta_l \mathbb{P}_l \Big)}=\sum_{l} \alpha_l^{k} \beta_l = O(1),
\end{equation*}
concluding the proof. $\blacksquare$ 

\subsection{Proof of Proposition 1}
\textit{Proof:}

First, consider a scrambler that lacks symmetries. According to Theorem 1, the only observables that avoid exponential concentration are those proportional to the identity.  This implies that symmetries are a necessary condition for efficiently extracting a non-trivial reservoir response, which is essential for properly solving the task of interest, as well as any other task. 

For a given $n$-qubit reservoir, we consider a unitary evolution with $n$ symmetry blocks:
\begin{equation*}
    U = \prod_{l=1}^n U_l,
\end{equation*}
where the Hilbert space on which $U_l$ non-trivially acts is generated by computational basis states 
containing exactly $l$ ones, making its action global in the set of qubits. It corresponds to the fact that $U$ is symmetric with respect to the operator $S=\sum_{i=1}^n \sigma^z_i$.

For the sake of definiteness, we analyze the reservoir evolution when a series of $|1\rangle \langle1|$ states is injected. Applying Lemma 1 and performing a direct calculation, we can determine the evolution of the coefficients $\alpha^k_l$, which define the mean of the distribution from which the reservoir state is sampled:
\begin{eqnarray}
\left\{
\begin{array}{l}
\alpha^{k+1}_n = \alpha^{k}_{n} + \frac{1}{n} \alpha^{k}_{n-1}, \\
\alpha^{k+1}_l = \frac{l}{n} \alpha^{k}_l + \left(1 - \frac{l-1}{n} \right) \alpha^{k}_{l-1}, \quad (0 < l < n), \\
\alpha^{k+1}_0 = 0.
\end{array}
\right.
\label{Eq:Alpha_Ev}
\end{eqnarray}

From Eq.~\eqref{Eq:Alpha_Ev}, we readily observe that the $\alpha_l^k$ coefficients converge exponentially fast, as a function of the time steps, to the values:
\begin{equation*}
    \alpha_n^{\infty} = 1; \quad \alpha_l^{\infty} = 0 \quad \forall l \neq n,
\end{equation*}
with a characteristic convergence time whose upper bound is $O(n^2)$.

In other words, the final reservoir state can be considered as sampled from a distribution whose mean is given by:
\begin{equation}
     \bigotimes_{i=1}^n |1\rangle\langle1|_i.
    \label{Eq:mean_state}
\end{equation}
However, all distribution samples must be compatible with the symmetry expectation values in Eq.~\eqref{Eq:Alpha_Ev}. From Lemma 1, we know that $\alpha^k_l = \Tr\{\mathbb{P}_l \, \rho_k^I \otimes \Tr_m\{\rho_k^R\}\}$, which implies that the reservoir state can only converge to the state defined in Eq.~\eqref{Eq:mean_state}. This property can also be verified by computing the variance of the distribution, following a similar procedure to the one used in the proof of Theorem 1, and confirming that it is equal to zero.

Similarly, for a series of $|0\rangle \langle0|$ states, the reservoir state will converge to:
\begin{equation*}
    \bigotimes_{i=1}^n |0\rangle\langle0|_i.
\end{equation*}
Consequently, measuring $\sigma_i^z$ for any system qubit immediately allows for discriminating the injected time series, proving Proposition 1. $\blacksquare$

\section{Additional numerical results}

\subsection{Numerical verification of the Echo state property}
\begin{figure}[h]
    \centering  \includegraphics[width=0.5\linewidth, keepaspectratio]{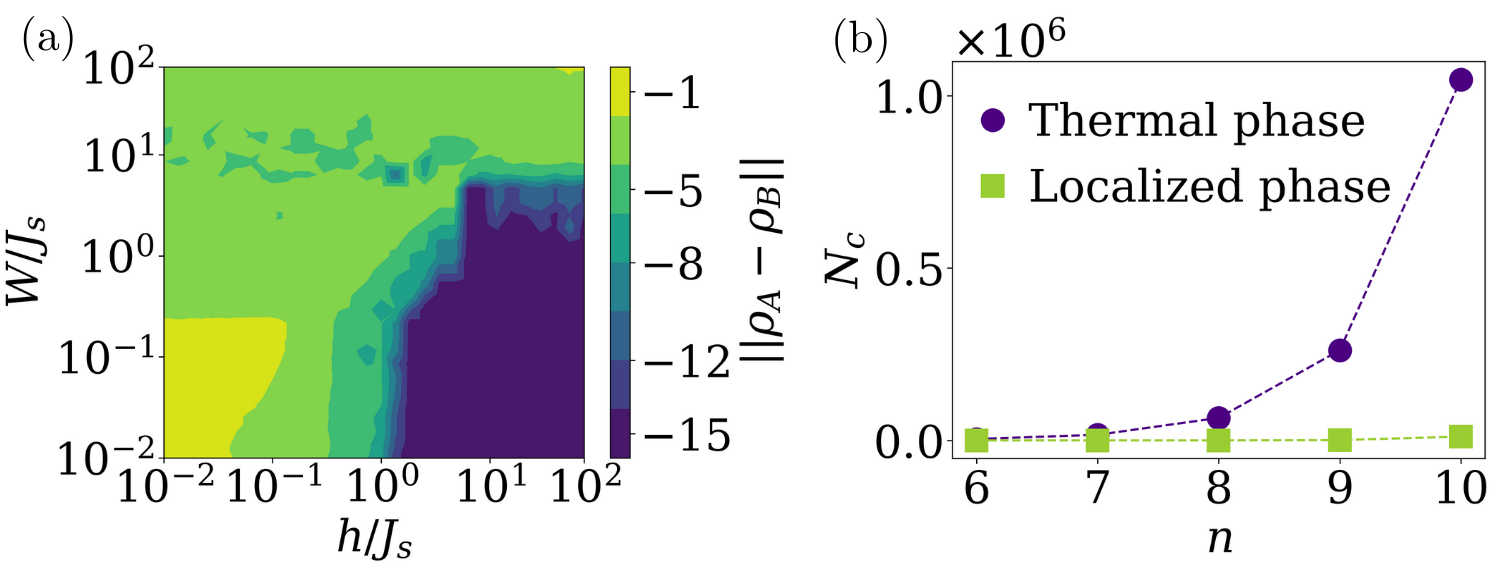}
    \caption{(a) Convergence of a 7-qubit reservoir state after the injection of 500 random inputs into two random initial conditions. (b) Number of observables that converged after the injection of 1000 inputs, as a function of the number of qubits. The results have been averaged over 100 random realizations, and possible statistical errors have been found to be negligible.}
    \label{fig:echo_state}
\end{figure}

Numerical verification of the echo state property for the fully transverse-field Ising model described in the main text. In Fig.~\ref{fig:echo_state} (a), we injected $500$ random inputs into two random initial conditions and we calculated the trace distance of the two final states, denoted as $\rho_A$ and $\rho_B$. We considered a $7$-qubit system, varying the hyperparameter values, and averaging over $100$ realizations of Hamiltonians and input series. 

Moreover, once the Hamiltonian phases were identified, we computed the number of degrees of freedom that properly converged, say $N_c$, fixing a threshold. In particular, similarly to the previous analysis, we injected $1000$ random inputs to a pair of random initial states, considering the cases of thermal ($W$=$10^{-1}$, $h$=$10^1$) and many-body localized phases ($W$= $10^2$, $h$=$10^1$). Averaging over $100$ realizations as in the previous case, we determined $N_c$ as the number elements $\Re\{(\rho_A-\rho_B)_{i,j}\}$ and $\Im\{(\rho_A-\rho_B)_{i,j}\}$ whose absolute value is below $10^{-10}$. This analysis is justified by the fact that in the limit of an infinite input sequence, the echo state property reads: $\rho_A$$-$$\rho_B$=$0$. In Fig.~\ref{fig:echo_state} (b), we can clearly observe that, in the thermal phase, $N_c$ tends to exponentially increase with the system size, while, in the localized one, we do not observe a significant change.

\end{document}